\documentclass[aps,prl]{revtex4}
\usepackage{graphicx}
\begin{document}
\title{Elucidation of the disulfide folding pathway of hirudin by a
 topology-based approach}
\date{\today}

\author{{\Large C. Micheletti$^{1}$,  V. De
Filippis$^2$, A. Maritan$^1$ and F. Seno$^3$,}\\
$\null$\\
$^1$International School for Advanced Studies, via Beirut 2-4, 34014
Trieste, Italy,  INFM and The Abdus Salam Centre for Theoretical Physics\\
$^2$ Dipartimento di Scienze Farmaceutiche and CRIBI Biotechnology
Centre, Via Marzolo 5, University of Padova, 35131 Italy \\
$^3$ INFM-Dipartimento di Fisica `G.
Galilei',  Via Marzolo 8, Universit\`a di Padova,  35131 Padova, Italy}

\begin{abstract}
A theoretical model for the folding of proteins containing disulfide
bonds is introduced. The model exploits the knowledge of the native
state to favour the progressive establishment of native
interactions. At variance with traditional approaches based on native
topology, not all native bonds are treated in the same way; in
particular, a suitable energy term is introduced to account for the
special strength of disulfide bonds (irrespective of whether they are
native or not) as well as their ability to undergo intra-molecular
reshuffling. The model thus possesses the minimal ingredients
necessary to investigated the much debated issue of whether the
re-folding process occurs through partially structured intermediates
with native or non-native disulfide bonds. This strategy is applied to
a context of particular interest, the re-folding process of Hirudin, a
thrombin-specific protease inhibitor, for which conflicting folding
pathways have been proposed. We show that the only two parameters in
the model (temperature and disulfide strength) can be tuned to
reproduce well a set of experimental transitions between species with
different number of formed disulfide.  This model is then used to
provide a characterisation of the folding process and a detailed
description of the species involved in the rate-limiting step of
Hirudin refolding.
\end{abstract}

\maketitle

\section{Introduction}

The characterisation of the folding pathway of proteins is one of the
fundamental problems in molecular biology and is under an increasing
scientific attention due to the continuous advancements in
experimental and theoretical biochemistry. After the work of
Anfinsen\cite{anfinsen}, who demonstrated that ribonuclease unfolds
and refolds reversibly into its native (active) three-dimensional
structure, it has generally been accepted that the primary sequence
usually contains sufficient information to direct the complete folding
process. What typically remains elusive to experimental and
theoretical investigations is the pathway of this spontaneous process
and the mechanisms that govern it.

A considerable progress in this direction is possible for proteins
containing native disulfide bonds.  The formation of disulfide bonds
during the folding process can be controlled experimentally through
the use of an appropriate thioldisulfide redox couple and thiol
quenching agent\cite{creighton,Cre92,Weis91}.  By these means the
regeneration process can be halted, the intermediate species can be
trapped, isolated and characterised. Historically, one of the most
investigated proteins containing disulfide bonds has been the bovine
pancreatic trypsin inhibitor (BPTI). Starting with the work of
Creighton \cite{Cre75} a series of crucial studies have accumulated a
wealth of evidence in favour of the existence of partially structured
intermediates along the protein folding pathway.  Despite these
efforts, the detailed characterisation of the intermediates turned out
to be a delicate experimental matter and there is still not a
universal agreement on whether intermediates contain native or
non-native disulfide bonds\cite{Weis91,Weis92,Cre92} and if there
exists more than one pathway\cite{Scher87,Weis91}. In this context,
the use of theoretical and computational tools
\cite{Cam95,Cam95b,AS00,TKD,CKS02} has been extremely useful in
complementing the experimental findings with a more precise
characterisation of the folding pathway, albeit obtained for models
that greatly simplify the complexity of the real system.

In this paper we propose a theoretical scheme to study the folding of
proteins with disulfide bonds by suitably exploiting the (known)
protein native structure. At a general level our strategy falls in the
class of approaches that build on the importance of the native state
topology in steering the folding process\cite{Go}, that is in bringing
into contact pairs of amino-acid that are found in interaction in the
native state. In the past few years an increasing number of
experimental and theoretical studies have confirmed the utility of
these approaches in the characterisation of various aspects of protein
folding processes
\cite{Miche99b,ME99,AB99,GF99,M00,Bak2000,CNO00,memb,hiv,cieplak,ZK99,SNGB98,Plotkin01}.
In the present work we try to generalise this strategy by adding a
suitable treatment of disulfide bonds accounting both for their
strength as well as their capability to undergo intra-molecular
reshuffling.

In order to validate this approach we investigate the folding process
of the N-terminal core domain of hirudin HM2 from the blood-sucking
leech Hirudinaria manillensis \cite{Scac93}. Hirudins are the most
potent and specific inhibitors of thrombin (a key enzyme in blood
coagulation) identified so far, and they are currently used as
effective anticoagulants. Hirudin HM2 is composed of a compact
N-terminal domain (residues 1-47) stabilised by three disulfides
(Cys6-Cys14, Cys16-Cys28, Cys22-Cys37) and a highly flexible,
negatively charged C-terminal tail. Structural studies conducted on
several leech-derived disulfide-rich small proteins (i.e., hirudin,
decorsin, and antistasin) reveal that although these proteins display
negligible sequence similarity and different function they share a
common disulfide topology and 3D fold \cite{Krez94}, suggesting that
leeches use the same protein scaffold but different binding epitopes
to affect hemostasis. Notably, it has been found that these leech
antihemostatic proteins possess a T-knot scaffold closely similar to
that observed for other unrelated proteins, including b-transforming
growth factors, wheat germ agglutinins and snake venom toxins
\cite{Lin95,Bol98}. In this view, the results of our study may have
relevant and more general implication on the elucidation of the
folding pathway(s) of other protein systems.

The first experimental attempts to identify the folding pathway of
Hirudin date back to the studies of Otto and Seckler who argued that
Hirudin could be a viable alternative to BPTI and showed that it was
experimentally feasible to obtain and follow its unfolding/refolding
processes\cite{OS91}.  As for the case of BPTI, also Hirudin has been
studied in different experiments \cite{OS91,Chat93,Def95,Scher97}
leading to alternative formulations of its folding pathway
\cite{Chat93,Scher97}. In particular when dissolved atmospheric oxygen
was used as oxidising agent, the folding process appeared to occur
first through the establishment of three non-native (scrambled)
disulfides and later through their slow rearrangement into the native
bonding pattern. On the contrary, no evidence for the importance of
these fully-oxidised disordered intermediates was found in the
experiments of Thannauser {\em et al.} carried out using oxidised
DL-ditiothreitol (DTT$^{ox}$).

These alternative views prompted the present investigation of the
Hirudin pathway within a topology-based model. To ascertain that,
despite its simplified nature, the model was suitable to characterize
the main aspects of the folding process we have first tuned and
validated it against the set of experimental measurements provided in
ref. \cite{Scher97}. Based on the success of this comparison we have
undertaken a detailed characterisation of the folding process by
monitoring quantities inaccessible in previous experiments.   Our
study confirms the experimental evidence of ref. \cite{Scher97}
suggesting that, under a certain set of experimental conditions, the
rate-limiting step of the refolding process involves a transition from
a species with two disulfides to ones with the three native
disulfides\cite{Scher97}. A detailed analysis of the numerical
dynamical trajectories further reveals the precise order of formation
of the disulfide bonds.

\section{Theory and results}

\begin{figure}[htbp]
\includegraphics[width=2.9in]{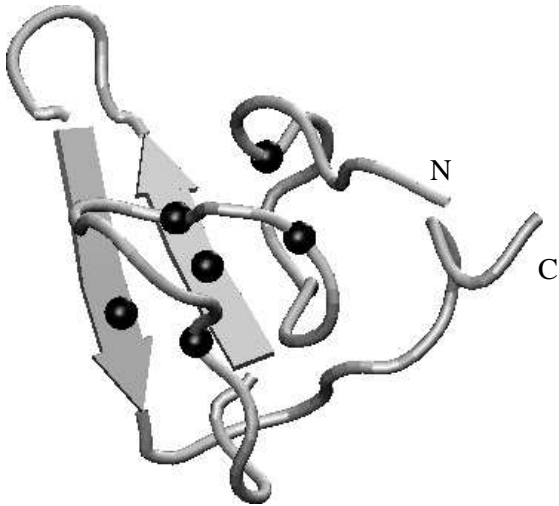}
\caption{Native structure of the 1-47 fragment of hirudin. The six
cysteine residues, 6, 14, 16, 22, 28, 37 have been highlighted.}
\label{fig:hirudin}
\end{figure}

The starting point of our analysis is the 1-47 fragment of
Hirudin\cite{Def95} resolved by NMR \cite{Nic97} shown in
Fig. \ref{fig:hirudin} (see Methods and Materials Section). The
fragment under consideration contains three disulfide bonds between
residues 6-14, 16-28, 22-37 and differs from the whole protein because
it lacks the 18-residues long tail; this tail plays an important role
in the biological activity of the protein. However, it is highly
mobile due to the virtual absence on any non-covalent contacts with
the 1-47 globular fragment.  For this reason we neglected the Hirudin
tail in our numerical characterisation of the folding process.

The effective energy scoring function that we adopted belongs to the
class of topology-based Hamiltonians \cite{Go,GA81}. The knowledge of
the above mentioned native structure, $\Gamma^N$, is exploited to
construct an effective energy function that admits $\Gamma^N$ as the
lowest energy state. The simplest form of such energy function is as
follows:

\begin{equation}
E(\Gamma) = - \sum \Delta_{ij}(\Gamma^N) \cdot \Delta_{ij}(\Gamma) \ ,
\label{eqn:simple}
\end{equation}

\noindent where $E(\Gamma)$ is the energy of a trial conformation
$\Gamma$ and $\Delta(\Gamma)$ is its contact matrix, whose elements
are 1 [0] if amino acids $i$ and $j$ are [not] in interaction. A
standard criterion based on $C_{\alpha}$ or $C_{\beta}$ distance of
pairs of amino acids is used to decide whether two amino acids are
interacting.

In the simple case of eqn. (\ref{eqn:simple}) the energy minimum is
attained when all native bonds are established. Such native
interactions are weighted equally. This is a simplification that can
be justified when the effective amino acid interactions in the protein
are of comparable strength and has the advantage of keeping the model
transparent by avoiding the use of imperfect energy parametrisation
\cite{maiorov:92,thomas:96:pnas,stabloc,chang}.

It is important to notice that the equal weighting of native contacts
does not imply that, in a thermalised ensemble, they are formed with
equal probability. In fact, it is the complex interplay of energy and
structural entropy that dictates the most probable routes to the
native state from an unfolded conformation as well as the presence of
rate-limiting steps due to the establishment of crucial sets of native
contacts\cite{hiv}.

In the present study, the equal weighting of native contacts does not
appear to be a good starting point since one has to account for the
very energetic disulfide bridges that can occur between pairs of
Cys. For this reason, in place of (\ref{eqn:simple}) we adopt a
scoring function consisting of two terms (see also section Methods
and Materials):

\begin{equation}
{\cal H} = V_{n-ss} + \mu \,V_{ss} \ .
\label{eqn:ham}
\end{equation}
\noindent
$V_{n-ss}$ enforces some general constraints on the peptide chain
geometry and promotes the formation of native contacts between pairs
of amino acids other than Cys-Cys. These contacts are weighted in the
same manner. The second term, $V_{ss}$ rewards the formation of
disulfide bonds between pairs of cysteines. It is important to stress
that the formation of disulfide bonds is allowed among any pair of
Cys, not only the native ones. By doing so we can investigate the
extent to which species with one or more non-native disulfides are
present and if they influence the dynamics toward the native state,
as proposed in recent experiments.

The strength of the disulfide bonds relative to other non-covalent
interactions is controlled by the parameter $\mu$. This parameter
should not be regarded as a relative measure of the ``bare'', i.e.
in vacuum, disulfide strength. In fact, since our model does not treat
the solvent explicitly, $\mu$ captures the effective strength of
disulfide bonds in the presence of water and any other appropriate
reducing/oxidising agent.
For this reason the value of $\mu$ and also that of the heat bath
temperature, $T$, have to be chosen in a suitable way so to reproduce
as accurately as possible the conditions of a given experiment.

In this study we shall focus on the set of experiments carried out by
Thannhauser {\em et al.}  \cite{Scher97} where hirudin was refolded in
the presence of various concentrations of $DTT^{ox}$.  We shall first
show that there exists a well-defined region in the $\mu-T$ parameter
space where the rates of conversion between species with different
numbers of disulfides match well those observed in experiments. This
is an important fact since it proves that, despite its simple form,
the energy scoring function of eqn. (\ref{eqn:ham}) can indeed be used
to characterize the folding process obtaining the correct quantitative
experimental picture. Based on such stringent validation of our
strategy, we shall then monitor quantities that are inaccessible in
current experiments and thus propose a vivid and detailed
picture for the hirudin refolding process.

We want to stress the fact that the possibility to reproducing the
results of Thannhauser {\em et al.}  \cite{Scher97} for a suitable choice of the
$\mu-T$ parameters support the hypothesis that our model is general
enough to study any folding process involving the formation of
disulfide bonds.

\subsection{Comparison with experimental rates}

\begin{figure}[htbp]
\includegraphics[width=2.9in]{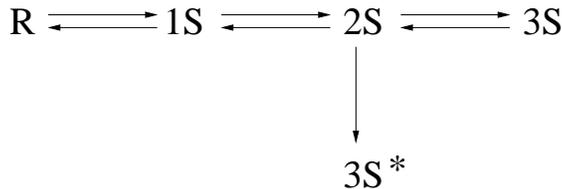}
\caption{Best fit model of Thannhauser {\em et al.} \cite{Scher97} to
the hirudin refolding experiments. Species with 0, 1 and 2 disulfides
are denoted as $R$, $1S$ and $2S$, respectively. The species denoted
as $3S^*$ contains the three native disulfides. The remainder of the
ensemble of structures with three disulfides is therefore
denoted as $3S$.}
\label{fig:model1}
\end{figure}

The experimental benchmark for our model is provided by a series of
hirudin refolding experiments carried out by Thannhauser {\em et al.}
\cite{Scher97} under various concentrations of $DTT^{ox}$. By using
several combined techniques it was established that the refolding
process occurs under the reaction shown in Fig. \ref{fig:model1} where
$R$, $1S$ and $2S$ denote respectively the species with 0, 1 and 2
disulfides (either native or non-native). A certain species with three
disulfides, denoted as $3S^*$ was seen to convert with extreme
rapidity to native hirudin and was, therefore, identified with a
native arrangement of the three disulfide contacts. The remainder of
the ensemble of structures with three disulfides (i.e. with at least
two non-native disulfides) is therefore denoted as $3S$.  The
experimental characterisation of Thannhauser {\em et al} provides the
rates for the individual reactions of the above model for a few
choices of the initial concentrations of $DTT^{ox}$ and the reduced
protein species. Our first goal was to see if, for suitably chosen
values of $T$ and $\mu$ it was possible to reproduce such rates.

Several values of $T$ and $\mu$ were considered and, for each of them,
we measured the rates of conversion between the $R$, $1S$, $2S$, $3S$
and $3S^*$ species by using a Monte Carlo technique. This was done by
keeping track of how many transitions out of any given species
occurred and how many of them ended up in each of the other
species. To ensure that the obtained dynamics could be interpreted as
a (coarse grained) viable dynamical trajectory \cite{Sokal} only tiny
local distortions of the peptide chain were attempted in the Monte
Carlo moves. During such coarse-grained trajectories the
formation, breaking or reshuffling of disulfide bonds obeys a set of
constraints dictated by the chemistry of the disulfide bonds and the
thiol-disulfide coupling. In particular no Cys residue is allowed to
participate to more than one disulfide; the
number of disulfide bonds can decrease or increase by at most one unit
at each time step, respectively when an existing disulfide is broken
or through the establishment of a new disulfide between two previously
unbonded Cys residues. Intramolecular rearrangements where one bond is
broken in favour of a new one (thus preserving the total number of
disulfides) are also allowed under the requirement that the new bond
involves one of the cysteines of the broken disulfide.

A particular care is necessary to ensure that the Monte Carlo dynamics
subject to these constraints does not violate detailed balance. The
difficulties arise from the fact that an elementary MC distortion of
distinct starting configurations may result in structures that are
compatible with a different number of allowed disulfide bonding patterns.

To overcome this problem we have proceeded as follows: to a newly
generated conformation $\Gamma$ we randomly associate one of the 15
possible pairing patterns of the three disulfides. If the
proposed bonding pattern of the new structure violates the previous
criteria, the structure is rejected and time is
incremented. Otherwise, the energy function is recalculated and the
structure is rejected or accepted with the usual Metropolis criterion
(see Methods for further details). In this way, detailed balance is
obviously satisfied, since the same number of bonding patterns is
proposed for each structure. The downside is that one encounters
frequent Metropolis rejections due to the ``blind'' proposal of
bonding patterns.

For each of the explored values of $T$ and $\mu$ we have measured the
correlation between the experimental rates and those obtained
numerically.  As a measure of the degree of correlation between these
two quantities we used the non-parametric Kendall analysis
\cite{NR}. This statistical tool allows to establish if there is a
relationship between two sets of data and how statistically
significant it is.  Being based on the comparative ranking of
corresponding data in the two sets, the Kendall analysis does not rely
on any pre-assigned parametric dependence (e.g. linear) between the
two quantities. For this reason, it is regarded as a very robust
measure of correlation and appears particularly appropriate in this
context where the measured rates (both experimental and numerical)
span several orders of magnitude \cite{NR}.

\begin{figure}[htbp]
\includegraphics[width=2.9in]{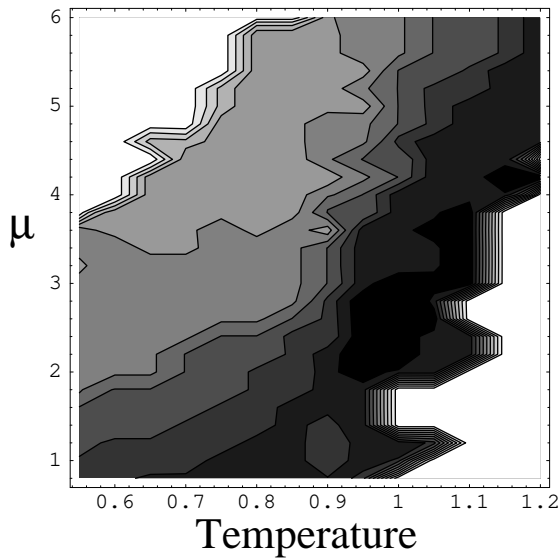}
\caption{Contour and density plot of the Kendall correlation across
the $T - \mu$ plane. The white regions correspond to the minimum
values of the correlations were observed ($\tau \approx 0$). The
highest correlation, $\tau=0.81$, was observed in the dark area
located around $\mu=2.8$, $T=1.0$.}
\label{fig:correl}
\end{figure}

Our findings are summarised in Fig. \ref{fig:correl} where we have
shown the contour and density plot of the Kendall correlation
coefficient, $\tau$, against the rates pertaining to the experimental
conditions of Fig. 5a in ref. \cite{Scher97}. These effective
first-order like rates where obtained starting from the best-fit
experimental rates of Table 2 in ref \cite{Scher97} multiplied by the
asymptotic concentration of DTT$^{ox}$ or DTT$^{red}$ measured under
the given experimental conditions (see Fig. 6 in the same
reference). A darker/lighter shadow in Fig. \ref{fig:correl} denotes
the presence of a higher/lower degree of correlation.  It is apparent
that the experimental data are well reproduced in the neighbourhood of
$\mu=2.8$, $T=1.0$ for which the highest correlation $\tau = 0.81 $ is
observed. To ascertain the statistical significance of observing such
correlation, we have computed the probability to observe, by pure
chance a correlation larger than the observed one. It turns out that
this probability (double-sided) is equal to $p=0.01$, which testifies
the statistical significance of the observed correlation.  This
establishes that there is a strong monotonic relation between the
experimental and theoretical rates.

\begin{figure}[htbp]
(a)\includegraphics[width=2.7in]{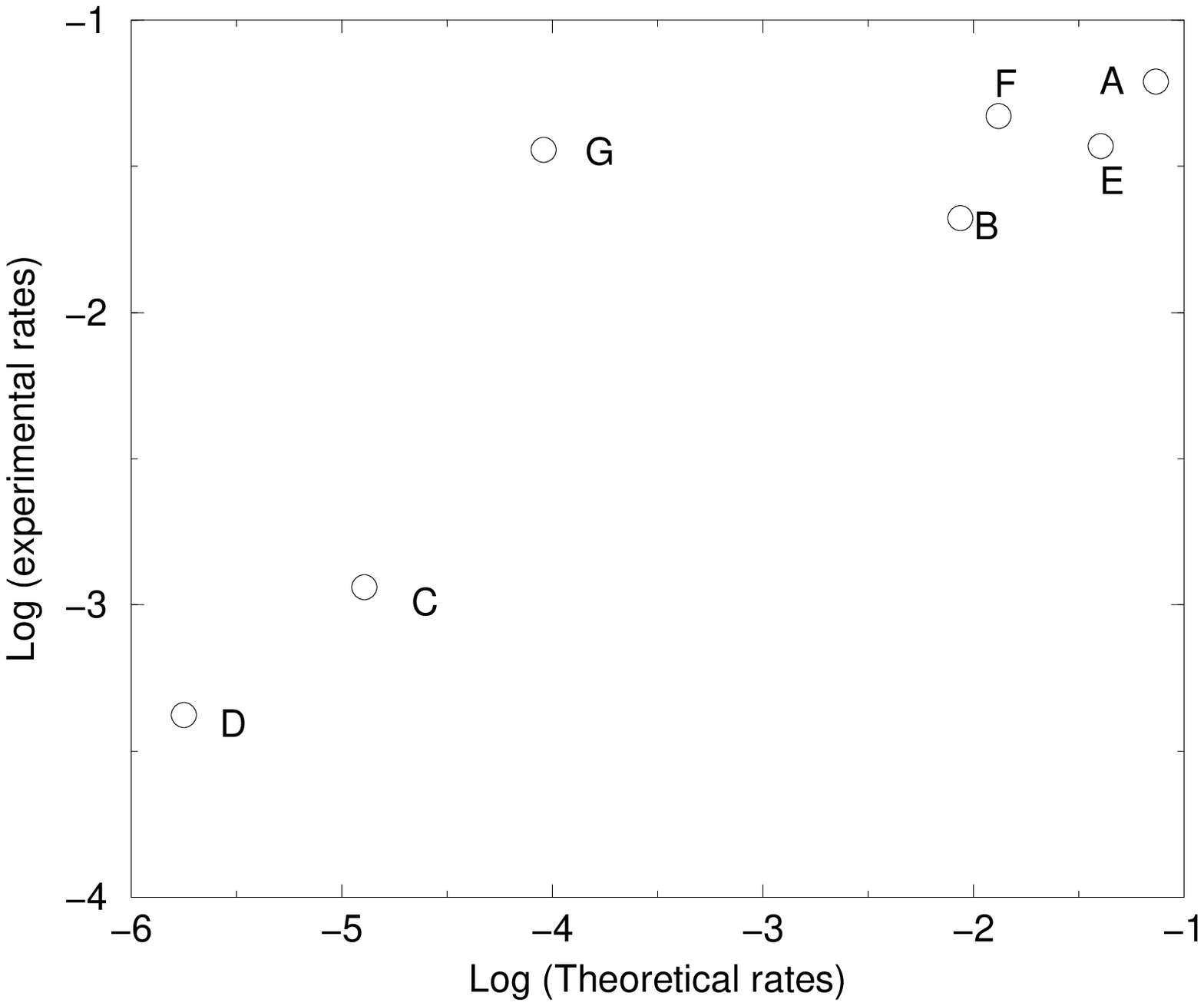}
(b)\includegraphics[width=2.7in]{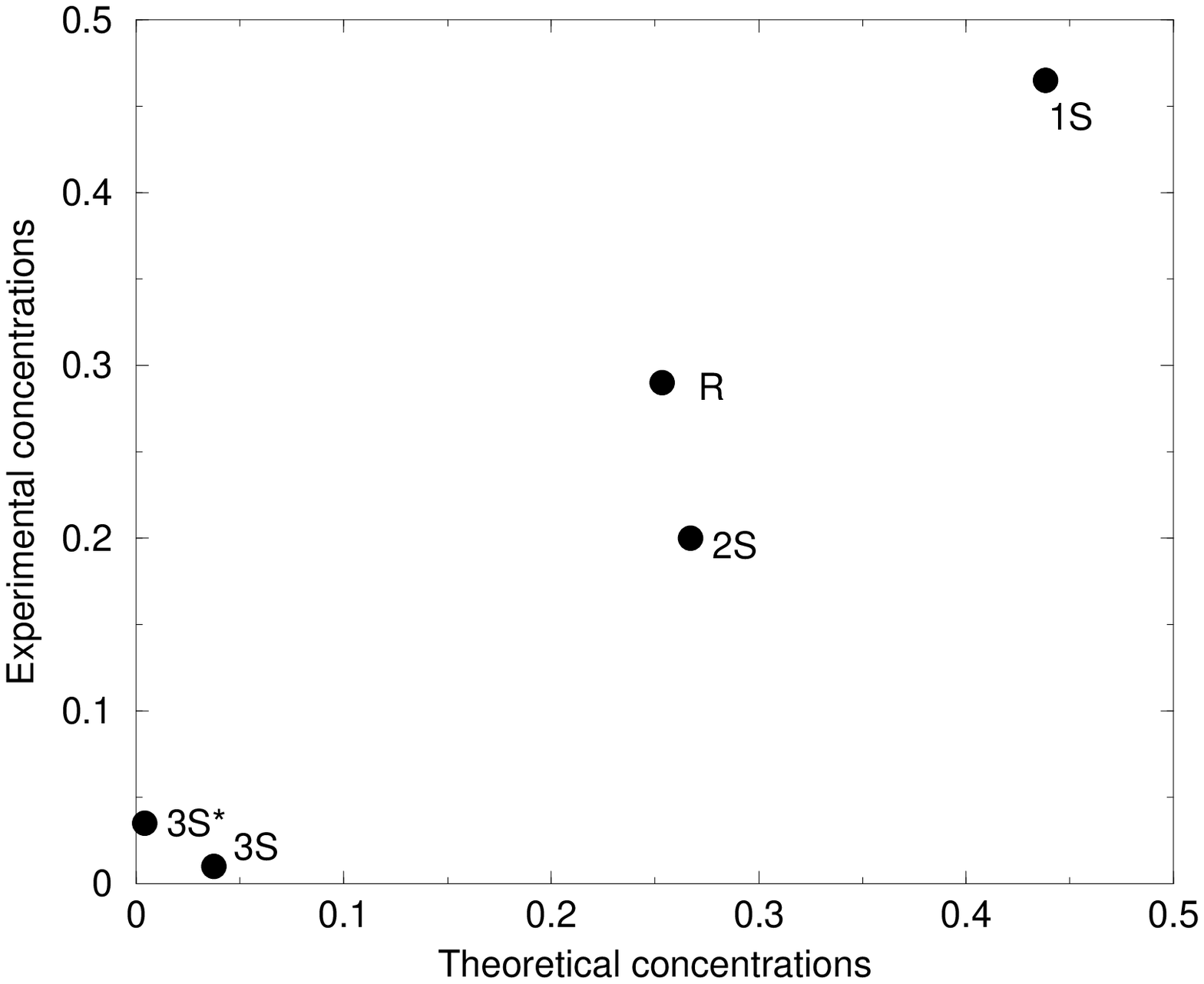}
\caption{(a) Linear correlation between the log of experimental rates of
Figs. 5a ref. \protect\cite{Scher97} and those obtained through the present
numerical calculation for $\mu=2.8$, $T=1.0$. The experimental rates
are expressed in min$^{-1}$. The points in the plot
correspond to the following transitions A: $R\to 1S$, B: $1S \to
2S$, C: $2S \to 3S$, D: $2S \to 3S^*$, E: $1S \to R$, F: $2S \to 1S$,
G: $3S \to 2S$. (b) Scatter plot of the equilibrium fraction of
the various species obtained in the Monte Carlo trajectory and in the
experiment. The experimental concentrations were extracted from Fig. 5a
of ref\protect\cite{Scher97} for $t=500$ s.}
\label{fig:opt}
\end{figure}

A scatter plot of the logarithm of experimental rates versus the
numerically obtained ones is provided in Fig. \ref{fig:opt} where the
good interdependence of the data can be visually inspected. If one
were able to model the folding process with detailed and accurate
energy functions one would expect an equality of the theoretical and
experimental rates, a part from a time conversion factor. In our case,
due to the simplicity of our model we do not observe such dependence
but encounter, instead, another simple linear relationship between the
logarithms of both sets of rates. This is visible in the linear fit of
Figure \ref{fig:correl}; the corresponding correlation coefficient is
$r=0.88$ . Its statistical significance (two-sided) over the set of 7
data points is 4\% which is compatible with the significance of the
more general (and robust) Kendall correlation. These values indicate
that the correlation is highly significant from the statistical point
of view.

 A further validation can be carried out by comparing the asymptotic
concentration of the various species observed in the experiment and in
an equilibrated MC trajectory. In case of a perfect correlation
between the experimental and numerical rates, this further check would
be redundant. In this context, where the correlation is not perfect (a
significant discrepancy is seen for the transition between the $3S$
and $2S$ species) this validation is useful to ascertain whether the
differences in the rates result in significantly different equilibrium
conditions. The plot of Fig. \ref{fig:opt}b reveals that the
asymptotic concentrations are in good accord and hence by setting
$T=1$ and $\mu=2.8$ one can be confident that the MC trajectory is
compatible both the dynamical and equilibrium properties of the
experimental system.

\subsection{Thermodynamics}

\begin{figure}[htbp]
(a)\includegraphics[width=2.7in]{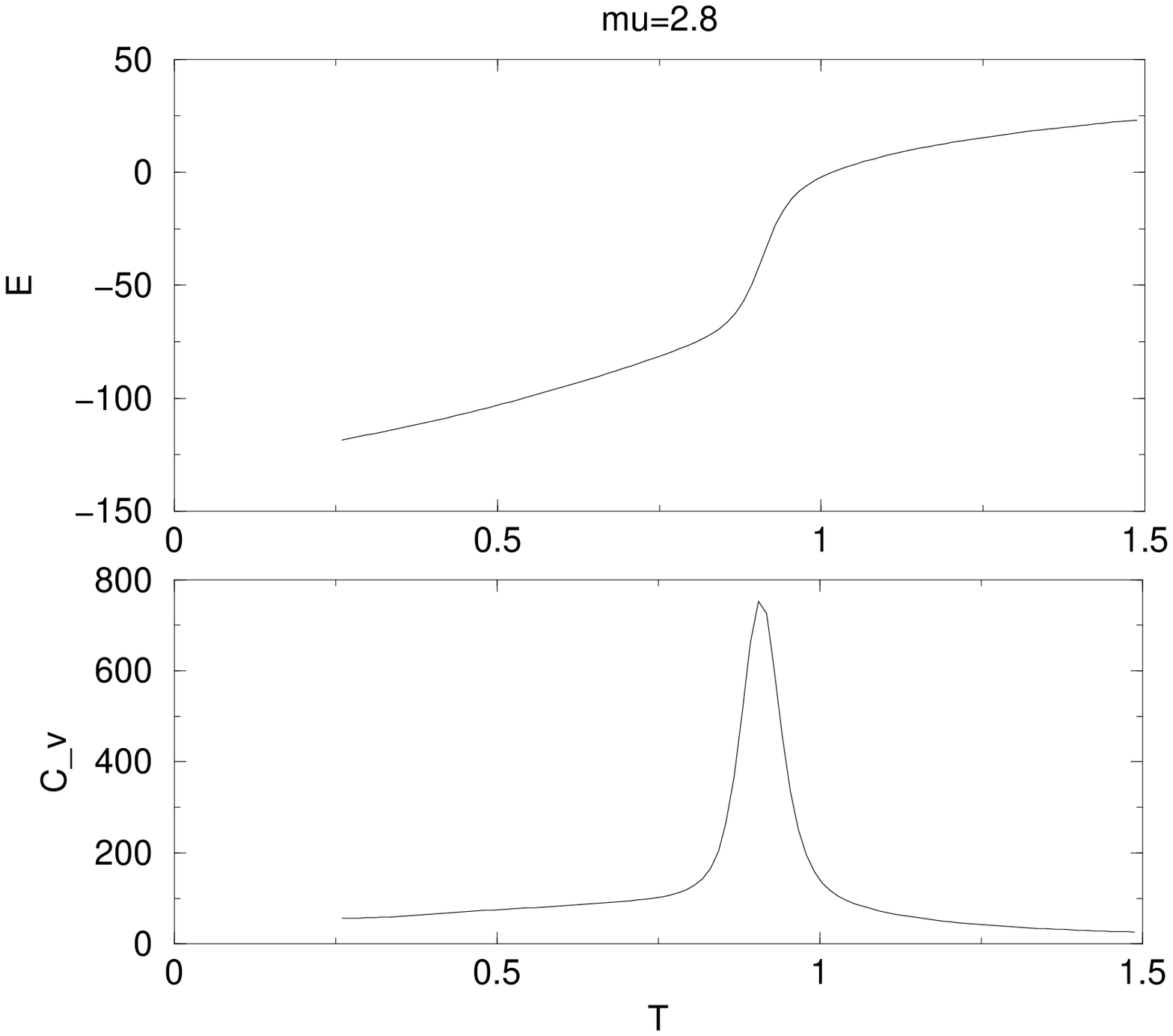}
(b)\includegraphics[width=2.7in]{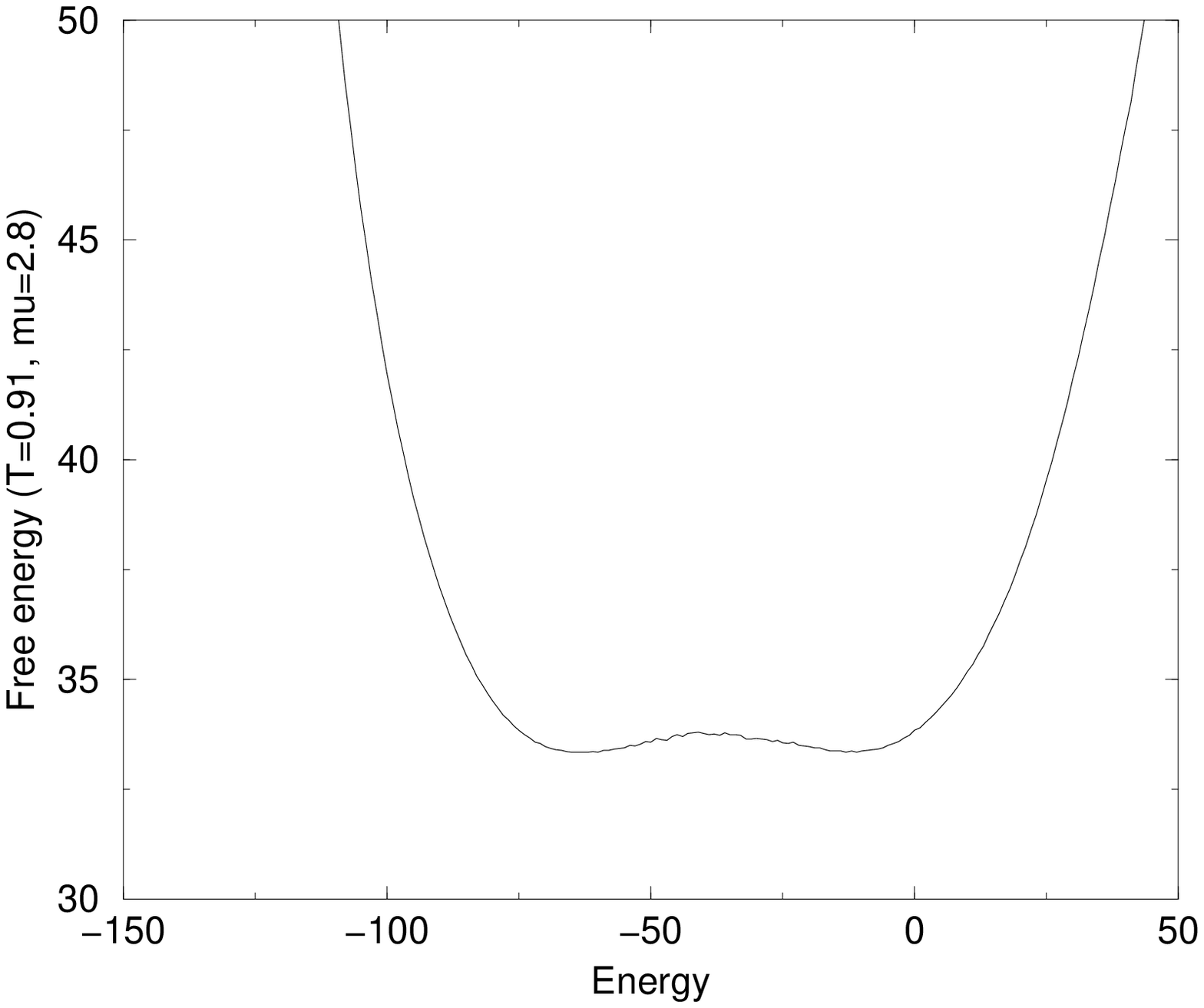}
\caption{(a) Total energy and specific heat for $\mu=2.8$.
(b) Free energy profile, as a function of the internal energy, at the
point $\mu=2.8$, $T=0.91$.}
\label{fig:encvmu5}
\end{figure}

It is interesting to analyse the thermodynamic behaviour of the system
at $\mu=2.8$ as a function of the heat bath temperature, $T$. By using
the standard, yet powerful method of histogram re-weighting (see
Methods) we have computed the average internal energy as a function of
$T$. The data, shown in Fig. \ref{fig:encvmu5}(a) indicate the presence
of a point of inflection for temperatures close to 1. The presence of
a transition at this temperature is further corroborated by the
behaviour of the specific heat $C_v$ which displays a clear peak at
$T=T_F \approx 0.91$, the folding temperature, and by the presence of
two minima in the free energy profile in Fig. \ref{fig:encvmu5}(b).

This peak is associated to the folding transition in the system and
its neatness suggests that the folding process has a two-state
character \cite{CKS02}. This is indeed confirmed by an
analysis of the free energy landscape at $T_F$ which exhibits two
minima as a function of $V_{n-ss}$ in correspondence of the unfolded
and folded states (data not shown). The special point $\mu=2.8$,
$T=1.0$, corresponding to the experimental conditions of
ref. \cite{Scher97} appears therefore to be located slightly above the
folding transition temperature (for $\mu=2.8$). This is entirely
consistent with the fact that the experimental conditions of
ref. \cite{Scher97}, Fig. 5a, do not particularly favour the formation
of the native state which, indeed, involves only a few percent of the
total equilibrium population.

\begin{figure}[htbp]
\includegraphics[width=2.9in]{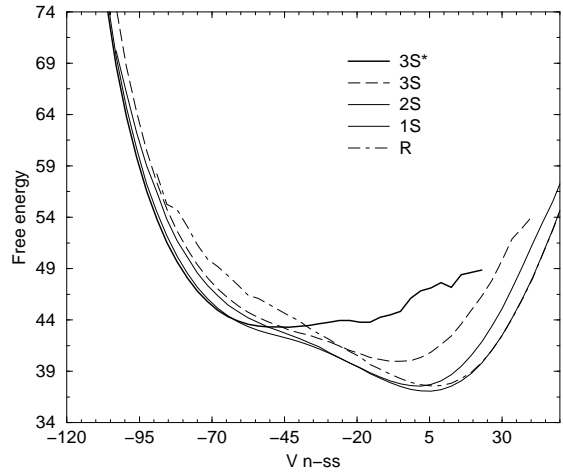}
\caption{Free energy profile for $T=1.0$ and $\mu=2.8$ of the $R$, $1S$,
  $2S$, $3S$ and $3S^*$ ensembles
as a function of $V^N_{n-ss}$.}
\label{fig:freemu5}
\end{figure}

We have portrayed in Fig. \ref{fig:freemu5} the free energy
landscape for the different reduced species as a function of
$V_{n-ss}$. It can be seen that the free energy profiles have minima
at different energy values according to the number of correctly formed
disulfides.  The $R$, $1S$, $2S$ and $3S$ species, in fact, display a
minimum near $E \approx 0$ which denotes an unfolded ensemble (see
Fig. \ref{fig:encvmu5}b). On the contrary, the species with three
native disulfides has a free-energy minimum for energies much closer to
the native state energy.

\begin{figure}
(a)\includegraphics[width=2.9in]{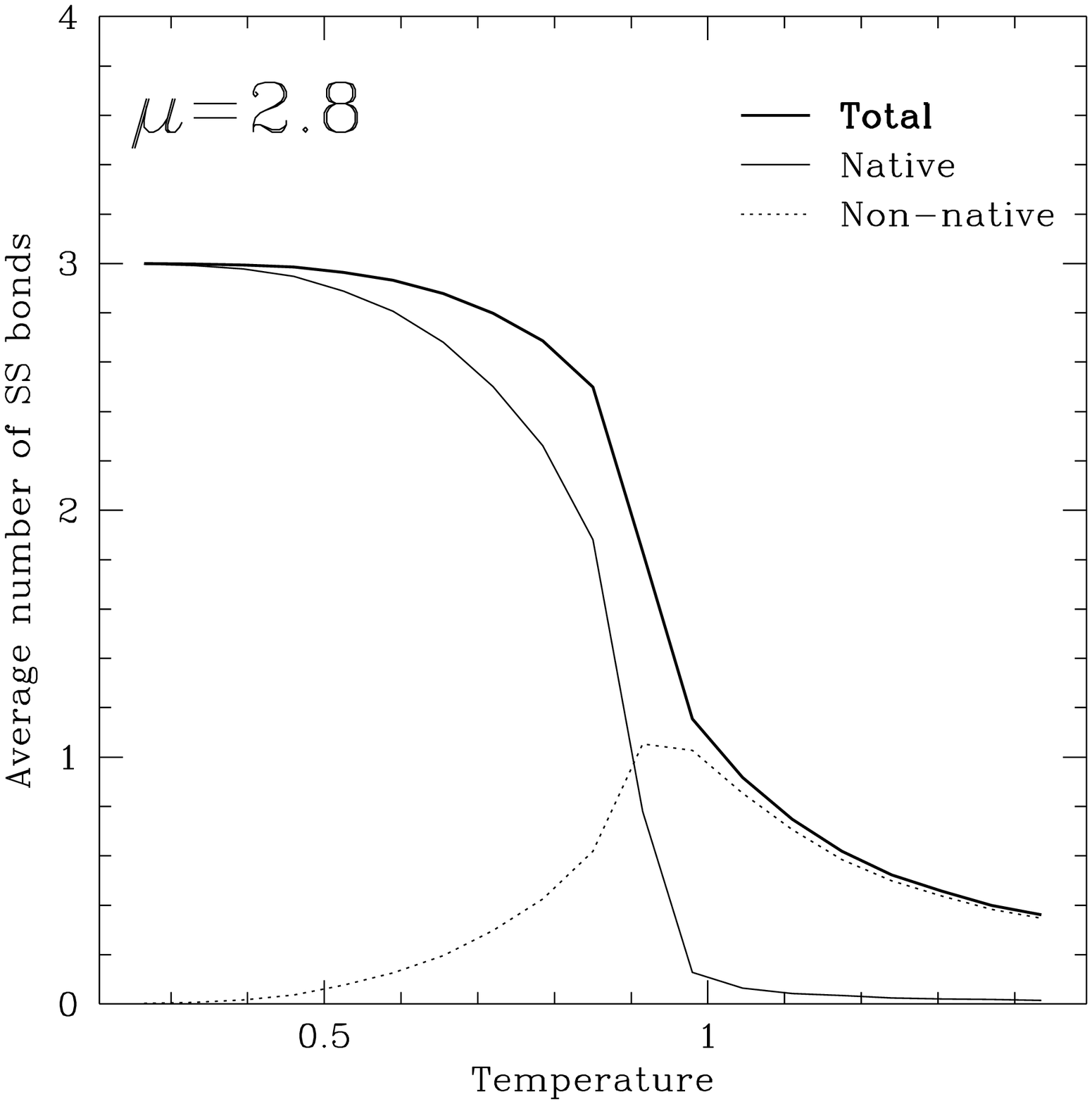}
(b)\includegraphics[width=2.9in]{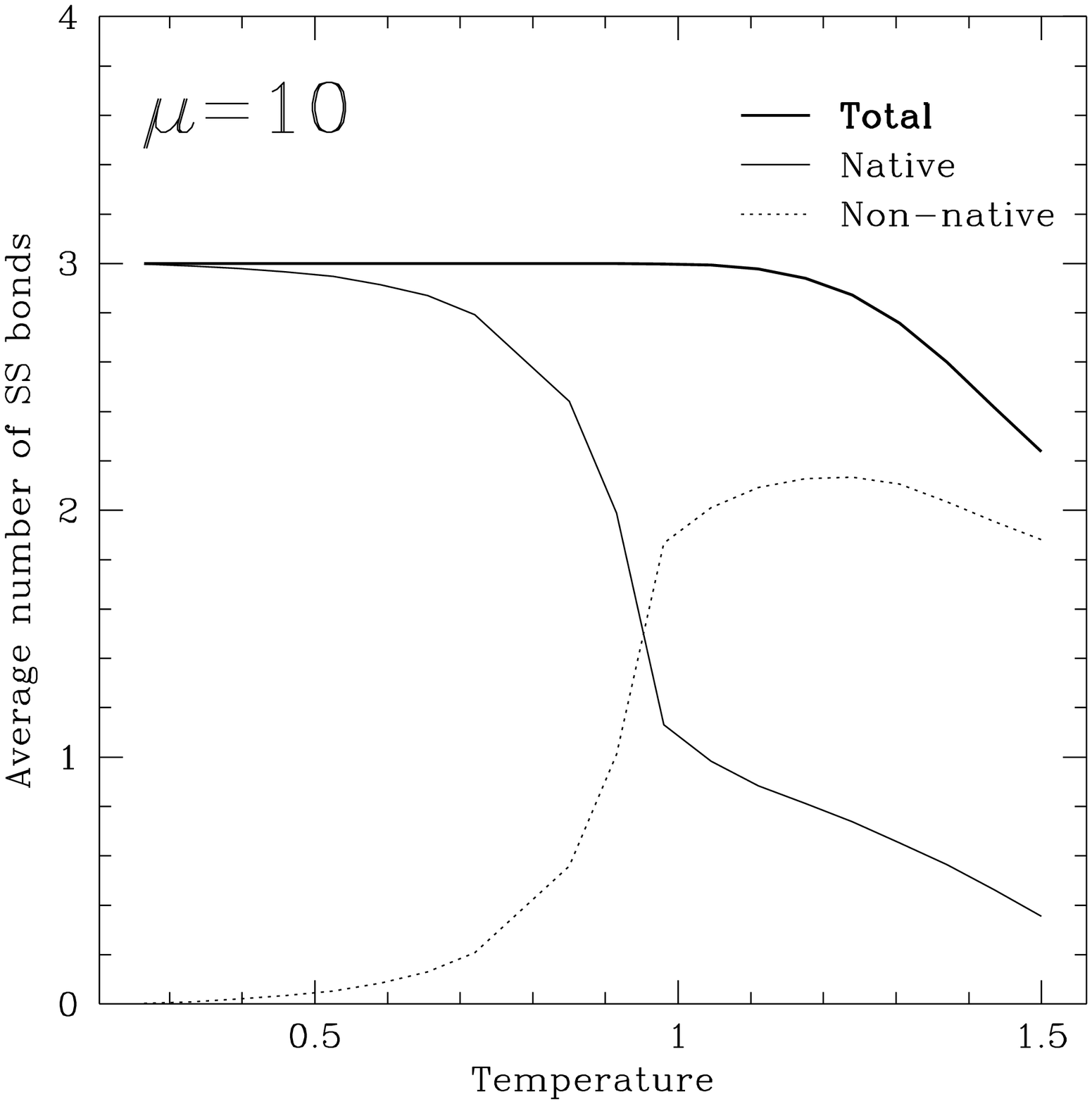}
\caption{Average number of total (thick continuous line), native
(continuous line) and non-native (dotted line) disulfide bonds for
(a)$\mu=2.8$ and (b)$\mu=10$.}
\label{fig:ss}
\end{figure}

 The relative high values of free energy associated to the $3S^*$
species reflect the particular experimental conditions reproduced here
where the asymptotic fraction of species with native disulfides is
low. Clearly, by lowering the temperature one favours the formation
of the native structure which is accompanied by an increase of the
concentration with native bonding patterns for the cysteines. This
effect is visible in Fig. \ref{fig:ss}a where the average fraction of
formed disulfide bonds is portrayed as a function of temperature. It
can be seen that above $T_F$ one has a significant formation of
non-native bonds, that are superseded by native ones below $T_F$.

An alternative picture to this one was put forward by Chatrenet and
Chang based on a hirudin refolding experiment carried out in
ref. \cite{Chat93}. Due to the much more oxidising conditions than the
one considered in ref. \cite{Scher97}, Chatrenet and Chang
\cite{Chat93} observed that the folding of hirudin occurred through the
formation of intermediates with three (typically non-native) formed
disulfides. Through a slow reshuffling process the disulfides would
then rearrange in the native pairing from which the native
conformation could be easily reached.

Our findings indicate that this alternative scenario could be probably
captured by our model by using larger values of the effective
disulfide strength $\mu$. This is consistent with the different
oxidising solvent conditions \cite{Scher97} adopted by Chatrenet and
Chang.

To investigate this regime we explored the value of $\mu= 10$.  It is
important to point out that the folding transition temperature
(identified through the location of the specific heat peak) is almost
insensitive to the disulfide strength in the range $0 \le \mu \le 10$.
By comparing the plots in Fig. \ref{fig:ss} it is then possible to see
that for the higher value of $\mu$ the total number of formed
disulfide bonds is higher than for $\mu=2.8$. This result,
accompanied by the fact that the total energy depends weakly
on $\mu$, confirms the intuition that, for large $\mu$'s the
disulfides are established at early stages of the folding process when
the rest of the protein is still unstructured. This ``greedy''
disulfide formation not only impacts on the total number of formed
disulfides but also on the relative fraction of correct (native
ones). As a result, a much higher fraction of wrong bonds is found at
any temperature, as noticeable in Fig. \ref{fig:ss}b. This picture
suggests that the large $\mu$ regime of our model may be compatible
with the scenario proposed by Chatrenet and Chang where the
rate-limiting step corresponded to the disentanglement of non-native
disulfides in intermediates states with three formed disulfide
bonds. However, for this alternative case the experimental rates are
not available and we cannot corroborate in a more quantitative way the
parallel between the experimental conditions of Chatrenet and Chang
and the ``large'' $\mu$ regime in our model.

\section{Folding pathways}

So far we have focused on the overall thermodynamic characterisation
of the refolding of hirudin, paying a particular attention to the
validation of our model against experimental data. Having established
that the detailed experimental reaction rates of ref. \cite{Scher97}
can be well reproduced we can confidently use our model to investigate
finer aspects of the refolding process.

\begin{figure}[htbp]
\includegraphics[width=2.9in]{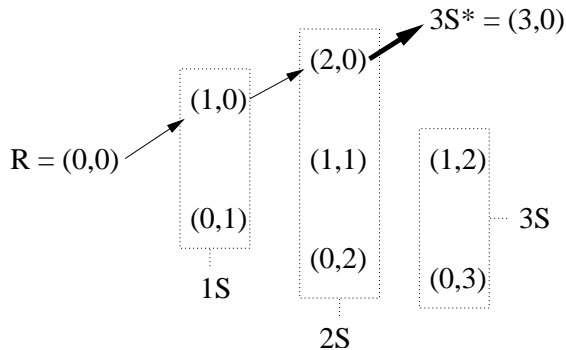}
\caption{The different states considered in our trajectory
analysis. The arrows indicate the most probable route from the reduced
ensemble toward the native arrangement of disulfides. The thickest
arrow denotes the rate-limiting step of the reaction.}
\label{fig:model2}
\end{figure}

We begin by examining the details of formation of the native
arrangement of disulfides. Our interest is to find whether the
formation of the $3S^*$ species is aided by the establishment of some
non-native disulfides that are later broken in favour of native
bondings.  To do so, instead of subdividing the structures according
to just the overall number of disulfides we indexed them with a pair
of numbers, $(n_c, n_w)$ denoting the number of correct (native)
disulfides, $n_c$ and wrong (non-native) ones, $n_w$. In this way the
fully reduced state, $R$, is indicated as (0,0) while the $3S^*$ state
corresponds to (3,0). Altogether there are nine possible states, as
indicated in Fig.\ref{fig:model2}.

We have generated several MC trajectories, each spanning about 20
million time steps (see sections Methods and Materials) at ($T=1.0$,
$\mu=2.8$) and then recorded the probability of occurrence of all
allowed transitions between the states of Fig. \ref{fig:model2}. The
typical dispersion on the measured rates over ten MC trajectories was
typically less than 1\% but augmented to about 10 \% for the few very
improbable transitions. Then we have considered all possible
productive routes taking from $R$ to $3S^*$ in a preassigned number of
transitions. The probability of occurrence for any such routes is
obtained by multiplying the individual probabilities of any of the
elementary steps.  By considering productive routes involving an
increasing number of transitions it can be established that the most
probable route (see Fig. \ref{fig:model2}) is $R \equiv (0,0) \to
(1,0) \to (2,0) \to (3,0) \equiv 3S^*$ which is more than ten times as
likely than the second ranking paths which is $R \to (0,1) \to (1,0)
\to (2,0) \to \equiv 3S^*$ . These findings seem robust against
variations of the parameters in our model. For example, even changing
the temperature to 0.8 (keping $\mu=2.8$) so that the formation of
native species is strongly favoured (see Fig. \ref{fig:ss}a) it is
found that the top ranking routes taking from the reduced state to the
native one are the same as above. Interestingly, the relative weight ratio of
the top productive routes is analogous to the one encountered for
$T=1.0$; on the other hand the fact that the native formation is
highly favoured at $T=0.8$ is reflected in an increase, by several
orders of magnitude, of the weight of the productive routes.

We complete the present section by identifying the rate limiting steps
of the folding process. In a sequential reaction the rate limiting
step is straightforwardly identified as the slowest one. In the
present scheme such simple analysis cannot be carried out due to the
presence of several alternative routes that can take to the native
state.  An objective and convenient way to identify the rate limiting
step in such a situation is to identify the reaction whose rate change
affects the most the production of the species of interest, in this
case $3S^*$.  Therefore, by using the measured rates for the
transitions between the states of Fig. \ref{fig:model2} we have
integrated the associated master equations starting from a fully
reduced population. We have then changed by 10 \% each of the rates
and identified the time at which the concentration of $3S^*$ crosses a
pre-assigned threshold value.  We found that, almost independently of
the threshold value, the most sensitive step among all the allowed
ones was $(2,0) \to 3S^*$ which is the last step of the most probable
route leading to $3S^*$. In principle this may have not been the case,
especially in the presence of equally important paths leading to
$3S^*$. The fact that the most probable routes include the rate
limiting steps confirms the existence of a well defined succession of
events taking to the native state.

As mentioned before, based on experimental dynamical plots,
Thannhauser {\em et al.}  \cite{Scher97} had determined that the rate
limiting step was the $2S \to 3S^*$ one. In their study it was not
possible to characterise by direct means whether the transition to the
$3S^*$ state occurred from a $2S$ state comprising only native
cysteine bonds, although this was reputed to be the most likely
scenario.  Our picture thus fully supports the experimental results
concerning the $2S \to 3S^*$ rate-limiting step, but also adds novel
insight in the process by indicating explicitly that the most crucial
step involves a particular $2S$ species, namely one with two native
disulfides. In the remainder of this section we shall further
characterise the folding process by identifying which of the native
disulfides are formed in the most probable routes (or rate-limiting
steps). This is done by following a series of individual dynamical
trajectories where the native conformation is reached starting from a
reduced state.

\subsection{Folding trajectories}

The Monte-Carlo procedure allows for a detailed study of the folding
pathways and for a systematic analysis of the specific order of
formation of disulfide bonds. By setting $\mu=2.8$ the protein is
initially thermalised at an high temperature ($T \sim 10$) where, in
our model, the disulfide bonds are completely reduced and then
suddenly quenched to $T = 1.0$.  For each kinetic trajectory the order
of formation of the disulfide bonds was stored and a statistical
analysis accomplished by comparing several runs.  In particular we
have recorded the exact type of disulfide bridges forming the
different species ($1S$,$2S$) and from their relative concentration we
have inferred the pathway.

The six Cys residues of the fully reduced protein, 6, 14, 16, 22, 28, 37
are equally likely to participate in forming the initial disulfide
bond. Although in the early folding stages the contact $(14-16)$
appears quite frequently due to the short sequence distance between
the amino-acids, after the molecule has equilibrated through a series
of rapid internal disulfide interchange reactions, only 5 of the 15
possible $1S$ states exist in significant quantities.  They are:
(6-14), (6-16), (16-22), (28-37) that are respectively present in the
following equilibrium concentrations: $7 \%$ $6 \%$ $12\%$ and $26
\%$. It is interesting to notice that the most common bond among the
$1S$ ensemble is a non-native one, (28-37), while the only native bond
that is present in significant quantities is (6-14).

Of the 45 possible $2S$ states, seven occur in significant quantities.
Two of them, (6-14;16-28) present in the $1 \%$ of the cases and
(6-14,22-37) present in the $0.7 \%$ of the cases, are formed by
native bonds.  Other two, (6-14;16-22) and (6-14,28-37) present with
concentration of $2 \%$ and $5 \%$ respectively, have a native and a
non-native contact. Whereas the last three, (6-16,28-37) with a
frequency of $9 \%$, (14-22,28-37) with a frequency of $7 \%$ and
(16-22,28-37) with a frequency of $7 \%$, are formed by non-native
disulfide bonds. The relatively high concentration of species with
non-native states, which are not present in the most probable
productive routes expected for the folding (see the previous section),
reflect the particular experimental conditions reproduced here
(induced by the choice of $T$ and $\mu$) in which the formation of the
native state is not particularly favoured (see Fig. \ref{fig:opt}b).

\begin{figure}[htbp]
\includegraphics[width=6.0in]{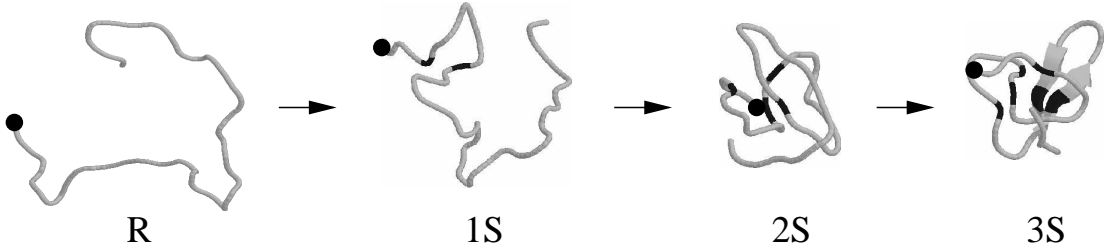}
\caption{Snapshots of one of the possible folding trajectories. The
 1S state involves the native contact (6-14), while
the  2S species involves the native bonds (6-14) and
 (16-28). A black sphere is used to distinguish the N-terminus while
cysteines involved in disulfide bonds are highlighted as dark
segments.}
\label{fig:pathway}
\end{figure}

The analysis of the folding pathway reported in the previous section
and the statistical analysis of the dynamical productive pathway can
be summarised in the following re-folding picture: the reduced
proteins forms first the native contact $(6-14)$ and consequently
either the states (6-14;16-28) or (6-14,28-37) almost with the same
probability.  The folding proceeds with the formation of the last
disulfide bond. This conversion is relatively slow in agreement with
the finding of the previous section on the rate limiting step of the
reaction.  The presence of conformation with three scrambled disulfide
bonds turns out not to be statistically significant also from this
kinetic analysis.  A graphical representation of the most probable
trajectory is shown in Fig.  \ref{fig:pathway}.

The fact that the majority of structures sampled in the quenching
process involve non-native bonding patterns is not in contradiction
with the findings of the previous section, where the most probable
{\em productive} route was shown to be free of non-native bonds.  In
fact, the high concentration of structures with non-native bonds does
not automatically imply that they contribute significantly to the
refolding ``flux'' towards the native state.  On the contrary, species
involving native bonds even if they do not accumulate to high
equilibrium concentration, appear to take part to the most efficient
routes leading to the native state.

\section{Methods and Materials}

\subsection{Structure of Hirudin}

The structure of the synthetic analogues of fragment 1-47 of hirudin
HM2 in the free state was modelled on the NMR solution structure of
natural fragment 1-47 \cite{Nic97}, almost super-imposable to that of
the corresponding segment (1-49) in intact hirudin HV1
variant\cite{Folkers89} (PDB code: 5HIR), showing $75 \%$ sequence
identity with hirudin HM2, or to that of HV1 fragment 1-51 \cite{Szy92}
(PDB code: 1HIC). The best-representative model in the NMR ensemble
was selected using the program OLDERADO \cite{Kell97} available
on-line at the site http://neon.chem.le.ac.uk.

\subsection{The Model}

In our model a generic conformation $\Gamma\{ \vec{r}_i\}$ of the
protein is modelled as a self-avoiding chain of connected $C_{\alpha}$
atoms located at position $\vec{r}_i^{{\alpha}}$, where $i$ is the
amino-acid chain index (ranging from 1 to 47). Starting from the
$C_{\alpha}$ coordinates the peptide dihedral angles are calculated
and hence, following standard geometrical rules \cite{PL96,stabloc}
we construct an effective  $C_\beta$ centroid for all residues
with the exception of the two end residues and the seven Glycines (indices:
10,18,23,25,34,40,42).

The energy scoring function consists of two terms. The first one
incorporates a standard bias toward the formation of native
non-disulfide bonds but also disfavours the formation of non-native
bonds (except for disulfide ones) and penalises significant deviations
from the native dihedral angles. These details are known to improve
the cooperatively of the modelled folding process
\cite{chan90,Kaya1,Kaya2,Settanni}. The explicit expression of this
term, evaluate on a trial structure $\Gamma$ is given by:

\begin{eqnarray}
 &V_{n-ss}(\Gamma)=\nonumber \\
&V_0\  \sum^{'}_{i,j>i+2}
\Delta_{ij}(\Gamma^{N}) \left[
5 \left(
{\frac{r^N_{ij}}{r_{ij}}}
\right)^{12} \right] \nonumber \\
& -  6 \left(
{\frac{r^N_{ij}}{r_{ij}}}
\right)^{10} \\
&+  V_1 \ \sum^{'}_{i,j>i+2}  \left[
1-\Delta_{ij}(\Gamma^{N})
\right]
\left(
\frac{r^N_{ij}}{r_{ij}}
\right)^{12}
 \nonumber \\
&+ V_2 \left[ \sum_{i=2,46} (\theta_i - \theta_i^{N})^{2}
+\sum_{i=3,46} (\phi_i - \phi_i^{N})^{2} \right]\nonumber \\
&+ V_{constraints}
\end{eqnarray}

\noindent where $r_{ij}$ denotes the distance of the $i$th and $j$th
$C_{\alpha}$ atoms in the trial structure $\Gamma$; a superscript $N$
is used to denote analogous quantities pertaining to the native
structure. The contact map $\Delta$ is computed by considering as
threshold a distance of 8 \AA $\ $ between the $C_{\alpha}$ atoms.
The prime in the summation indicates that no contribution is
considered if both the amino-acids are Cys.  Finally, the angular term
has been constructed using the standard dihedral angles, $\theta$ and
$\phi$ \cite{PL96,stabloc}. The coefficients $V_0$,$V_1$ and $V_2$ are
used to control the strength of interactions and are set equal to 1, 5
and 1 energy units, respectively.   The last term in the
expression, $V_{constraints}$, is used to enforce a series of
knowledge-based constraints whose violation is penalised through an
``infinite'' energy penalty (that is through a rejection of the
violating conformation). The constraints are as follows: (1) the
distance between two consecutive $C_{\alpha}$ atoms must remain in the
interval 3.7-3.9 \AA, (2) the distance between two non consecutive
$C_{\alpha}$ atoms must be greater than 4 \AA, (3) the distance
between any two $C_{\beta}$ must be greater than 2 \AA $\ $ and (4)
the distance between any two $C_{\alpha}$ and $C_{\beta}$ centroids
must be greater than 2 \AA.

The second term of the Hamiltonian, $V_{ss}$ rewards the formation of
disulfides, irrespective of whether they are native or not.  As
explained in the next subsection, each proposed configuration,
described in terms of $C_\alpha$'s and $C_\beta$'s, comes with a set
of three putative bonds among the six cysteines. Of these putative
bonds only those among residues whose $C_\beta$'s are a separation smaller
than 5 \AA\ are considered to be effectively present and hence give a
contribution equal to $-\mu$ to the total energy.

\subsection{Monte-Carlo Method}

As mentioned in the text, we used Monte Carlo dynamics for studying
the folding process. At each Monte-Carlo step the current structure is
distorted through local deformations \cite{oligons} based on two
equally-probable moves: (a) {\em single-bead move:} a random
$alpha$-carbon is chosen and is displaced randomly by at most 1 \AA\
along each Cartesian direction, (b) {\em crankshaft move:} two sites,
$i$ and $j$, with sequence separation at most 6 are chosen and all the
sites between them are rotated around the axis joining $i$ and $j$ by
an angle chosen randomly in the interval $-\frac{\pi}{10} \leq \Omega
\leq \frac{\pi}{10}$. As mentioned before, besides this structural
rearrangement, at each Monte Carlo step we also associate a
randomly-chosen pairing pattern for the six cysteines so that each of
them is involved in a {\em putative} disulfide bond. These bonds are
termed putative because, to satisfy detailed balance, the pairing
assignment is done blindly that is without inspecting whether a given
pair of cysteines is at a distance compatible with the existence of a
disulfide bond. One then checks whether (irrespective of the pairing
distances) the proposed bonding pattern is compatible with the
undistorted configuration, i.e. if it respects the rules of section
2.1 about disulfide formation and thiol/disulfide coupling. If not the
configuration is rejected, the Monte Carlo clock is advanced and a new
distortion and bonding pattern is considered. Otherwise one proceeds
as in ordinary Metropolis schemes after having calculated the energy
of the proposed configuration. It is important to notice that this
latter step involves the inspection of the distance of the putatively
bonded cysteines to reward only those bonds that are geometrically
feasible.

The efficiency of the Monte-Carlo algorithm to study the
thermodynamics was enhanced by the multiple Markov-chain sampling
scheme\cite{TROW}, a method that has proved quite effective in
exploring the low temperature phase diagrams of proteins and
 interacting polymers. All the runs have been performed by
covering the temperature range of interest, $T=[0.2,1.5]$, with 20
Markov chains uniformly-spaced in temperature.

The data obtained in the multiple-Markov-chain runs at different
values of $T$ and $\mu$ were further processed through a generalised
multiple histogram technique inspired by the work of
ref. \cite{ferren}. This strategy allowed to reconstruct faithfully
the density of states (number of configurations) in the
multi-dimensional space of reaction coordinates constituted by
$V_{n-ss}$ and the number of correct (native) and wrong (non-native)
disulfides: $(n_c,n_w)$. The data from the different runs were
combined so to minimize the error propagation on the density of
states. The typical uncertainty of the reconstructed free-energy at
the values of $T$ and $\mu$ considered here is about 0.5 energy
unit (this estimate follows from the analysis of free energy
dispersion when half of the collected data is used).
By these means it was possible to obtain the total energy and specific
heat curves of Fig. \ref{fig:encvmu5} and the free energy profiles of
Fig. \ref{fig:freemu5}.

\section{Conclusions}

We have proposed a theoretical framework to model and study the
folding of proteins containing disulfide bonds. The approach is based
on the knowledge of the native state of a protein but contains an
appropriate term to account for the possibility that native or
non-native disulfide bonds can form.  The main advantage of the model
proposed here is its simplicity which allows for a detailed
description of all the folding pathways through the monitoring of the
correct/incorrect contact formation.

The model has been validated by investigating the debated re-folding
pathways of Hirudin which has been object of several experimental
studies. It was shown that there exists a region in our
two-dimensional parameter space where the rates of conversions between
different oxidised species are in good agreement with experimental
measurements \cite{Scher97}. Starting from this successful comparison
we have then attempted a detailed characterisation of the whole
folding process.

At a coarse-grained level our results is consistent with the scenario
described by Thannhauser {\em et al.} \cite{Scher97} suggesting that
the rate limiting step turns out to be $2S \to 3S^{*}$.  Our
approach, that  allow for a  precise identification of the formed
contacts, shows clearly that   the $2S$ state appears to involve native
intermediates, possibility that was  reputed
 as the most probable situation in
ref. \cite{Scher97} although experimentally it was impossible
to show it directly.  The analysis of several folding trajectories
also allowed the identification of the most probable folding route and
the typical associated succession of formation of disulfide
bridges. Furthermore, the thermodynamics of the system was elucidated
by using statistical mechanical techniques to reconstruct the free
energy profiles for the whole system and also for the different
oxidised species.

Due to its simplicity, the proposed model, cannot capture those
aspects of the folding process that result from the delicate interplay
of amino acid specific interactions. The successful comparison of the
theoretical predictions for hirudin with the experimental findings
suggests that also, for disulfide containing proteins, suitable
topology-based models can be profitably used to elucidate the folding
pathways, even in the presence of non-native intermediates. Thus,
the present approach, which is general and not specifically tailored
for hirudin, ought to be straightforwardly applicable in other
contexts providing a useful complement of experimental techniques in
the characterisation of the folding process in the presence of
disulfide bonds.

\section{Acknowledgements}

We thank F. Cecconi, G.L. Lattanzi and D. Marenduzzo for stimulating
discussions and suggestions. This work was supported by INFM,
FISR-MIUR 2001 and Murst Cofin 2001.

\end{document}